\documentclass[12pt  ]
  {article}

\RequirePackage{graphicx}

\newcommand{\OL}[1]{ \hspace{1pt}\overline{\hspace{-.5pt}#1
     \hspace{-1.5pt}}\hspace{1.5pt} }
\newcommand{\klmt}{\mbox{K\hspace{-7pt}KLM\hspace{-9pt}MT}\ }
\def\lsim{\stackrel{<}{{_\sim}}}

\begin{document}

\title{Cosmic Superstrings Revisited}

\author{Joseph Polchinski\\[6pt]
\it \normalsize{ Kavli Institute for Theoretical Physics}\\
\it \normalsize{University of California}\\
\it   \normalsize{Santa Barbara, CA 93106-4030}}

\maketitle

\begin{abstract}
It is possible that superstrings, as well as other one-dimensional branes, could have been produced in the early universe and then expanded to cosmic size today.   I discuss the conditions under which this will occur, and the signatures of these strings.  Such cosmic superstrings could be the brightest objects visible in gravitational wave astronomy, and might be distinguishable from gauge theory cosmic strings by their network properties.

Presented at the Mitchell Conference on Superstring Cosmology, GR17, Cosmo 04, and the 2004 APS-DPF Meeting.
\end{abstract}
\bigskip
%%%%%%%%%%%%%%%%%%%%%%%%%%%%%%%%%%%%%%%%%%%%
%% MAINMATTER
%%%%%%%%%%%%%%%%%%%%%%%%%%%%%%%%%%%%%%%%%%%%
\baselineskip 16pt

Seeing superstrings of cosmic size would be a spectacular way to verify string theory.  Witten considered this possibility in the context of perturbative string theory, and found that it was excluded for several reasons~\cite{Witten:1985fp}.  Perturbative strings have a tension close to the Planck scale, and so would produce inhomogeneities in the cosmic microwave background far larger than observed.  The scale of this tension also exceeds the upper bound on the energy scale of the inflationary vacuum, and so these strings could not have been produced after inflation, and any strings produced earlier would have been diluted beyond observation.  Ref.~\cite{Witten:1985fp} also identified instabilities that would prevent long strings from surviving on cosmic time scales.

In recent years we have understood that there are much more general possibilities for the geometry of the compact dimensions of string theory, including localized branes, and this allows the string tension to be much lower, anything between the Planck scale and the weak scale.  Also, we have found new kinds of extended object in string theory. Thus the question of cosmic superstrings (and branes) must be revisited, and this has been done beginning in refs.~\cite{Jones:2002cv, Sarangi:2002yt}.  A necessary set of conditions is:
\begin{enumerate}
\item
The strings must be {\it produced} after inflation.
\item
They must be {\it stable} on cosmological timescales.
\item
They must be {\it observable} in some way, but not already excluded.  
\end{enumerate}
Ref.~\cite{Witten:1985fp} thus showed that perturbative strings fail on all three counts.  If we do find models that satisfy these three conditions then there is one more that would also be important:
\begin{itemize}
\item[4.] Cosmic superstrings should be {\it distinguishable} from other kinds of cosmic string, in particular gauge theory solitons.
\end{itemize}
We will see that each of these issues is separately model-dependent, but that there are simple models, including the most fully-developed models of inflation in string theory, in which all these conditions are met.  

The list above provides an outline for this paper, after an introductory section which reviews the story of cosmic strings in grand unified field theories.

\setcounter{section}{-1}
\section{Cosmic String Review}

Cosmic strings might also arise as gauge theory solitons, and for some time these were a candidate for the source of the gravitational perturbations that produced the galaxies.  It is useful to give a brief summary here to set a context for the superstring discussion, but for a more complete discussion and references see the reviews~\cite{VilShell,Hindmarsh:1994re}; for a recent overview see ref.~\cite{Kibblenew}.

In any field theory with a broken $U(1)$ symmetry, there will be classical solutions that are extended in one dimension~\cite{Abrikosov:1956sx, Nielsen:1973cs}.  These are topological solitons: as one traverses a circle around the string core, the Higgs field winds around the manifold of vacua, which is also a circle.  The broken $U(1)$ can be either a global or a gauge symmetry.  In the global case the Goldstone boson field has a $1/r$ gradient at long distance, giving a logarithmic potential between strings.  In the gauge case this gradient is pure gauge and the physical fields all fall off exponentially.  Most of the discussion of cosmic strings has focused on gauge strings, though one should keep the global case in mind for later reference.

These solutions exist whenever there is a broken $U(1)$ symmetry, and whenever a $U(1)$ symmetry {\it becomes} broken during the evolution of the universe a network of
strings must actually form.  This is the Kibble argument~\cite{Kibble:1976sj}: the Higgs field starts at zero, and then rolls down to one of the vacua.  By causality, it cannot roll in the same direction everywhere in the universe, as it cannot be correlated on distances greater than the horizon scale (in practice, the correlation length is usually less than this, being the microphysical correlation length of the field theory).  Thus it chooses random directions in different places, and inevitably there will be some trapped winding, so that strings are left over at the end.  This is indeed what simulations show.  A fraction of the string is in the form of finite sized loops, and a fraction is in the form of infinite strings; the latter enter and leave the boundaries of the simulation volume no matter how large this is taken to be.  These populations are cleanly separated because the distribution of lengths of finite loops falls rapidly for long loops, giving a convergent integral for the total amount of string in loops.  Presumably the existence of the infinite strings is implied by the causality argument. They are important because they begin to stretch with the expansion of the universe, while the small loops quickly decay away.

In the subsequent evolution of the network, the important processes in addition to the expansion of the universe are collisions of strings and the emission of gravitational radiation.\footnote{For reasons of space I am focusing on strings whose only long-distance interactions are gravitational.  Global strings would also emit Goldstone bosons.  There are also superconducting strings~\cite{Witten:1984eb}, which have strong couplings to gauge fields.  These are perhaps less likely to arise in the superstring case, for reasons that I will explain.}  When two strings collide they can either pass through each other or they can reconnect (intercommute), as in figure~1.
\begin{figure}[t]
\begin{center}
\includegraphics[height=.13\textheight]{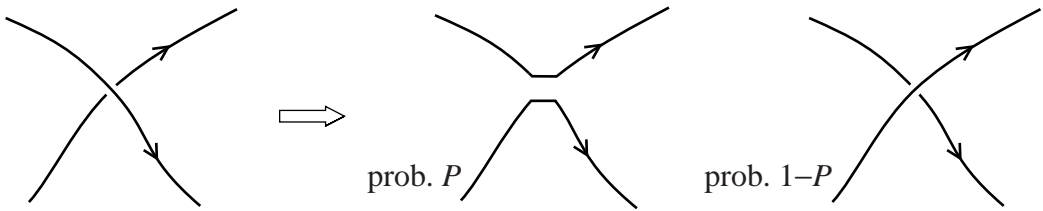}
\caption{When two strings of the same type collide, they either reconnect, with probability $P$, or pass through each other, with probability $1-P$.  For classical solitons the process is deterministic, and $P=1$ for the velocities relevant to the string network.}
\end{center}
\end{figure}
In an adiabatic collision, they will always reconnect, $P=1$, because reconnection allows the flux in the string core to take an energetically favorable shortcut.  Simulations show that this remains true at the moderately relativistic velocities that are present in string networks~\cite{Shellard:1987bv, Matzner, Moriarty:1988fx}.

During the radiation and matter dominated eras, the horizon distance grows linearly in $t$ while the comoving distance grows more slowly, as $t^{1/2}$ or $t^{2/3}$ respectively.  Thus as time goes on we see more of the universe, and if no decay processes were operating we would see more strings.  The relevant decay process is the combined effect of reconnection and gravitational radiation.  Reconnection turns long straight strings into long kinked strings and then allows loops of string to break off, and these loops decay by gravitational radiation.  Simulations show that this occurs at the maximum rate allowed by causality, so that the string network always looks the same as viewed at the horizon scale,  with a few dozen strings spanning the horizon volume and a gas of loops of various sizes.  This is known as the {\it scaling solution}.  The scaling solution is an attractor: if we start with too much string, the higher collision rate will reduce it, while if we start with too little then there will be few collisions until the amount of string per horizon volume approaches the scaling form.  It is possible that the strings have a short-distance structure that does not scale with the horizon size.

Since we are considering strings whose only interactions are gravitational, all of their effects are controlled by the dimensionless product of Newton's constant and the string tension, $G\mu$.  This is the string tension in Planck units, and it sets the size of the typical metric perturbation produced by a string.  For example, the geometry around a long straight string is conic, with a deficit angle $8\pi G\mu$.  

The string network produces inhomogeneities proportional to $G\mu$, and because of the scaling property of the string network these are scale invariant.  A value $G\mu \sim 10^{-5.5}$ would give scale-invariant perturbations of the right magnitude to produce the galaxies and CMB fluctuations.\footnote{I am going to quote values of $G\mu$ to the nearest half order of magnitude.  This is less precise than most numbers given in the literature, but it is all I will need and it  roughly reflects the uncertainties in the understanding of the string network.  To give more precise numbers would require a more detailed discussion, and so the reader should consult the references.  Incidentally, some of the bounds given here are stronger than those that I was aware of when this talk was given.} 
This was a viable theory for some time~\cite{Zeldovich:1980gh, Vilenkin:1981iu}, but it is now excluded.
For example, the CMB power spectrum is wrong: the actual spectrum shows a pattern of peaks and dips, whereas the spectrum from strings would be smooth.  There is a simple reason for this.  The fluctuations produced in inflation have a definite phase.  This phase is maintained from the end of inflation until the perturbations go nonlinear, and is imprinted as oscillations of the power spectrum.  Strings, on the other hand, each keep their own time, there is no common phase.  Fitting cosmological constant plus cold dark matter {\it plus} strings to the CMB power spectrum gives an upper limit $G\mu< 10^{-6}$~\cite{Pogosian:2003mz, Pogosian:2004ny}.  Beyond the power spectrum, strings will produce nongaussianities in the CMB.  Recent limits are somewhat stronger than those from the power spectrum, around $G\mu< 10^{-6.5}$~\cite{Jeong:2004ut,Lo}.

Another limit on $G\mu$ comes from pulsar timing.  Because the energy in the strings eventually goes into gravitational waves, strings produce a large stochastic gravitational wave background.  The classic reference on this subject~\cite{Kaspi:1994hp} quotes a limit on the energy density in stochastic gravitational waves, per logarithmic frequency range, as $\Omega_{\rm GW} < 1.2 \times 10^{-7}$, using $h^2 = 0.5$ for the Hubble parameter.  In the frequency range of interest one can estimate the stochastic background from a network of strings as $\Omega_{\rm GW} = 0.04 G\mu$~\cite{VilShell}, meaning that $G\mu < 10^{-5.5}$.  Different analyses of {\it the same} data, with different statistical methods, give a (controversial) bound a factor of 6 stronger~\cite{Thorsett:1996dr} and another a factor of 1.6 weaker~\cite{McHugh:1996hd}.  The limits from pulsars should increase rapidly with greater observation time, and recent work~\cite{Lommen:2002je} quotes
a bound 30 times stronger than~\cite{Kaspi:1994hp}, meaning that $G\mu < 10^{-7}$.  However, this paper again obtains much weaker limits using other methods and so this should not be regarded as a bound until there is agreement on the analysis.\footnote{I would like to thank E. Flanagan and H. Tye for communications on this point.}  
The bounds here are from gravitational waves with wavelengths comparable to the size of the emitting string loop, and do not include a potentially substantial enhancement due to high-frequency cusps; see section~4.

Thus far we have quoted upper bounds, but there are possible detections of strings via gravitational lensing.  A long string will produce a pair of images symmetric about an axis, very different from lensing by a point mass.  Such an event has been reported recently~\cite{Sazhin:2003cp, Sazhin:2004fv}.  The separation of around two arc-seconds corresponds to  $G\mu$ equal to $4 \times 10^{-7}$ times a geometric factor that is at least 1. This appears to exceed the upper limit from pulsars, unless there exist fewer strings than in the usual network simulations.  I have heard varying opinions on how seriously to take such an observation, as similar pairs in the past have turned out to coincidental.  There is further discussion in the review~\cite{Kibblenew}, which also discusses a possible time-dependent lens, as from an oscillating loop, with $G\mu \sim 10^{-7.5}$~\cite{Schild:2004uv}.

Network simulations show that the total energy density in the string network is of order $60 G\mu$ times the matter density (during the matter-dominated era), so the upper bound on $G\mu$ implies that the strings contribute only a small fraction to the total energy density.  In particular, the strings are not the dark matter, if they have the usual network properties.

\section{Production of Cosmic F- and D-Strings}

In order to discuss inflationary cosmology, one needs a fairly clear understanding of the scalar field dynamics.  This has been an area of recent progress in string theory, and a nice geometric idea has emerged for obtaining a slow-roll potential~\cite{Dvali:1998pa,Alexander:2001ks,Burgess:2001fx,Dvali:2001fw}.  That is, the early universe could have contained an extra brane-antibrane pair, separated in the compact directions.  The potential energy of these branes would drive inflation.  The inflaton is then the separation between the branes: this has a potential which is rather flat when the branes are separated and steepens as they approach, until at some point a field becomes tachyonic and the brane-antibrane annihilate rapidly.

If these branes are D-branes, then there is a $U(1)$ gauge symmetry on each of the brane and antibrane, and this $U(1)\times U(1)$ disappears when the branes annihilate.  One linear combination of the $U(1)$'s is Higged.  The Kibble argument then applies, so that a network of strings must be left over when the branes annilate~\cite{Jones:2002cv, Sarangi:2002yt}.  These are D-strings, as one can see by studying the conserved charges~\cite{Sen:1998tt, Witten:1998cd}.  More precisely, if the branes that annihilate are $D(3+k)$ branes, extended in the three large dimensions and wrapped on $k$ small dimensions, then the result is $D(1+k)$ branes that extend in one large dimension and are wrapped on the same small dimensions.  The simplest case is $k=0$, where D3/$\OL{\rm D3}$ annihilation produces D1-branes.

It is important that this process produces only strings, and not monopoles or domain walls~\cite{Jones:2002cv, Sarangi:2002yt}.  We have seen that the existence of strings is consistent with observation, provided that the scale of their tension is a few orders of magnitude below the string scale.  By contrast, zero- and two-dimensional defects are observationally ruled out because if they were produced then their energy density would quickly come to dominate the universe (unless they were removed by instabilities, inflation, or further symmetry breaking).  The point here is that $\Pi_1$ topology from the breaking of a $U(1)$ produces defects of codimension two, and the Kibble argument requires that the codimension be in the large directions: the small directions are in causal contact.\footnote{Ref.~\cite{Dvali:2003zj} gives arguments that the Kibble process cannot produce D1-branes; we disagree.  The argument in Sec.~IV of that paper is based on the energetics of the Kaluza-Klein RR fields.  These are massive and should be integrated out, and so cannot affect the phase transition (the point is that one is not producing real KK particles, rather these fields are just adjusting in response to the light fields).  Also, the model in Sec.~V of that paper has its $\Pi_1$ disorder partly in the compact directions, which is not the case for the D-strings under consideration~\cite{Jones:2002cv, Sarangi:2002yt}.}

The second linear combination of $U(1)$'s is confined.  We can think of confinement as dual Higgsing, by a magnetically charged field, and so we would expect that again the Kibble argument implies production of strings~\cite{Dvali:2003zj, Copeland:2003bj}.  These are simply the F-strings, the `fundamental' superstrings whose quantization defines the theory, at least perturbatively.

It is striking that what appears to be the most natural implementation of inflation in string theory produces strings and not dangerous defects, but we should now ask how generic this is.  Even these models are not, in their current form, completely natural: like all models of inflation they require tuning at the per cent level to give a sufficiently flat potential~\cite{Kachru:2003sx}.  There might well be other flat regions in the large potential energy landscape of string theory.  An optimistic sign is that there are arguments entirely independent of string theory to indicate that inflation terminates with a symmetry-breaking transition: this is known as hybrid inflation~\cite{Linde:1993cn}, and leads to efficient reheating as well as production of strings~\cite{Yokoyama:1989pa,Kofman:1995fi,Tkachev:1998dc}.  However, not every symmetry-breaking pattern produces strings.  For example, in strongly coupled heterotic string theory~\cite{Witten:1996mz}, there are M5-branes and M2-branes.  It would be natural to use the M5-branes, with two wrapped dimensions, to implement brane inflation.  Since the M2-branes have codimension three relative to the M5-branes, it will not be so easy to produce strings; this is currently under investigation.\footnote{ There has also been recent discussion of a more exotic symmetry-breaking pattern in D-brane inflation~\cite{Urrestilla:2004eh,Watari:2004xh,Dasgupta:2004dw}.  For other discussions of string production in brane inflation see refs.~\cite{Halyo:2004zj,Matsuda:2004bk,Matsuda:2004nx}}
It may still be that our vacuum is well-described by weakly coupled heterotic string theory.  In that case inflation, and cosmic strings, might simply arise from the low energy effective field theory~\cite{Witten:1985fp}.

In~\cite{Englert:1988th} it was proposed that cosmic fundamental strings could have been produced in a Hagedorn transition in the early universe.  The Hagedorn transition corresponds to the formation of strings of infinite length, and so some percolating strings would survive as the universe cooled below the Hagedorn temperature.  This idea has a simple realization in the warped models.\footnote{Similar ideas are being considered by A. Frey and R. Myers.}  We have noted that the effective string tension, and so the Hagedorn temperature, is different in different throats.  It is possible that after inflation a deeper throat reheats above its Hagedorn temperature, leading to string formation as the universe cools (a black hole horizon would then form at the bottom of the throat, corresponding to the Hawking-Page transition~\cite{Hawking:1982dh,Witten:1998zw}).
In fact, this is essentially equivalent to the Kibble mechanism.  The throat degrees of freedom have a dual gauge description, in terms of which the Hagedorn transition corresponds to deconfinement.  The transition to a confining phase as the universe cools is the electric-magnetic dual of a symmetry-breaking transition.  The strings produced in this way would necessarily have a lower tension than those produced directly in brane annihilation, because of the inefficiency of the extra thermal step.

\section{Stability}

Ref.~\cite{Witten:1985fp} identified two instabilities that would prevent superstrings from growing to cosmic size.  Actually, these same two instabilities exist for field theory soliton strings~\cite{Preskill:1992ck} --- one for global strings and the other for gauge strings --- so let us first discuss them in this context.

In the case of global strings, we have noted that the long-ranged Goldstone boson has gradient energy.  It does not have potential energy at long distance as long as the broken $U(1)$ symmetry is exact: the broken vacua are then exactly degenerate.  However, there are general arguments in string theory that there are no exact global symmetries~\cite{Banks:1988yz,Polchinski:1998rr}.  More generally, the no-hair theorems imply that black holes can destroy global charges, so in any theory of gravity these can not be exactly conserved.
Thus the degeneracy of the vacua will not be exact, and there will be a potential energy cost at long distance from the global string.  This takes the form of a domain wall, with energy proportional to its area, bounded by the string.  The wall exerts a transverse force on the string and forces it to collapse, as in figure~2.
\begin{figure}[t]
\begin{center}
\includegraphics[height=.13\textheight]{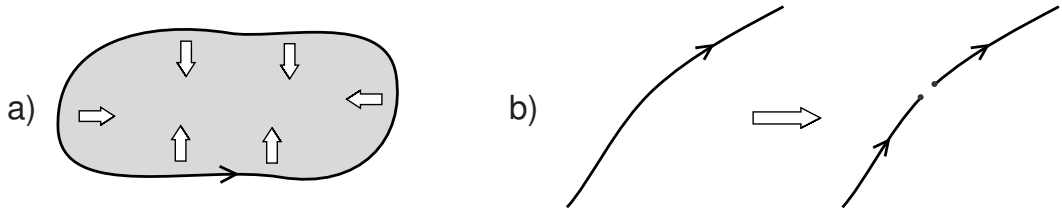}
\end{center}
\caption{Instabilities of macroscopic strings: a) Confinement by a domain wall.  b) Breakage.}
\end{figure}
  This is clear for a loop of string bounding a domain wall, but is less intuitive for a network with infinite strings.  One can picture the network, with the domain walls, as a complicated shape formed from strips whose width is the typical transverse separation between strings.  The timescale for the collapse of the strips, and the disappearance of the strings, is set by the width of the strips, not their (potentially unbounded) length~\cite{Vilenkin:1982ks}.

For gauge strings, the $U(1)$ symmetry is exact because it is a gauge symmetry, and all energies fall exponentially with distance from the string.  A magnetic flux runs along the core of the string, and it is the conservation of this flux that prevents the string from breaking.  However, in any unified theory one expects that there will be electric and magnetic sources for every flux~\cite{Polchinski:2003bq}, so that the string can break by creation of a monopole/antimonopole pair.  If this is possible, it will happen not only once but everywhere along the length of the string, and so the string breaks up into short segments rather thans growing to cosmic length.

These two instabilities are rather generic, but there are two ways that strings might evade them.  First, they might turn out to be slow on cosmic timescales because they are suppressed by large ratios of scales.  For monopole pair creation, for example, the classic Schwinger calculation gives the pair production rate as $e^{-\pi M^2/\mu}$, where $M$ is the monopole mass.  If the monopole mass is an order of magnitude larger than the scale of the string tension, then the decay will be slow on cosmological time scales.  One can think of a succession of symmetry breakings $SU(2) \to U(1) \to 1$.  There is no $U(1)$ at the beginning so there are no stable strings, but if the scale of the first symmetry breaking is higher than that of the second then there will be string solutions in the effective field theory describing the second breaking, and they can be long-lived.

Second, there is way that strings can be exactly stable.  Consider the Abrikosov flux tubes in an ordinary superconductor.  The Higgs field there is an electron pair and has charge $2e$, so the tube has total
flux $2\pi/2e$.  However, because there do exist singly charged electrons, Dirac quantization gives the minimum monopole charge as twice this, $2\pi/e$.  Thus a flux tube cannot end on a monopole, though two can; equivalently one can think of the monopole as a bead on a string, at which the flux reverses.  So the Abrikosov flux tubes (in an infinite system) are absolutely stable.  One can think of this in terms of an unbroken discrete gauge symmetry $(-1)^{Q/e}$, which acts as $-1$ on the electron and $+1$ on the BCS condensate.  As one circles the string, fields come back to themselves only up to this transformation.  Thus the string can be detected by an Aharonov-Bohm experiment at arbitrarily long distance, and so it can never just disappear.  We will refer to these absolutely stable strings as `discrete' strings.  Gauge strings without a discrete gauge charge are truly invisible at long distance, and so there is no obstacle to their breaking.

Now let us turn to string theory.  Consider first perturbative strings~\cite{Witten:1985fp}, compactified for example on a torus or a Calabi-Yau manifold.  The heterotic and type II strings are effectively global strings, because they couple to the long-distance form field $B_{\mu\nu}$.  The confining force is then produced by instantons, which are magnetic sources for these fields (thus the completeness of the magnetic sources~\cite{Polchinski:2003bq} again enters).  It seems difficult to suppress the confining force enough for the strings to survive to today --- in particular, QCD instantons contribute in the heterotic case and give a lower bound on the force.
The type I string couples to no form field, and of course it can break.  The time scale for the breaking is the string scale unless one tunes the string coupling essentially to zero.

The most fully developed model of inflation in string theory is the \klmt model~\cite{Kachru:2003sx}, in which inflation is due to a D3/$\OL{\rm D3}$ pair.  This model has F- and D1-strings, and these are effectively gauge strings.  In ten dimensions these strings couple to form fields, but there are no massless form fields in four dimensions; they are removed by the orientifold or F theory monodromy that the model requires.  Thus the strings are unstable to breakage, and this can occur in several ways.

First, the projection that removes the form field produces an oppositely oriented image string on the covering space of the compactification, and breakage occurs through a segment of the string annihilating with its image.  If the image is coincident with the string, breakage will be rapid.  If the image is not coincident, then the string must fluctuate to find its image.  At this point an important feature of the \klmt model enters, the warping of the compact dimensions.  That is, the metric is
\begin{equation}
ds^2 = e^{2\Delta(y)} \eta_{\mu\nu} dx^\mu dx^\nu + \ldots\ .
\end{equation}
The factor $e^{\Delta(y)}$ is a gravitational redshift which varies strongly as a function of the compact coordinates $y$.  Generally $e^{\Delta(y)}$ is around 1 over most of the compact space, but falls to much smaller values in a few small regions known as `throats.'  The effective tension of a string as seen by a four-dimensional physicist depends on its ten-dimensional tension $\mu_0$ and the local gravitational redshift at its position:
\begin{equation}
\mu = e^{2\Delta(y)} \mu_0 \ .  \label{eq:tenred}
\end{equation}
This is the Randall-Sundrum (RS) idea~\cite{Randall:1999ee}, that different four-dimensional scales arise from a single underlying scale through gravitational redshifting.  This means that the strings feel a strong potential with local minima in the throats.  The same is true for the inflationary D3/$\OL{\rm D3}$: these sit near the bottom of some throat, and the strings produced in their annihilation sit near the bottom of the same throat.  As we will discuss further in the next section, the depth of this throat is at least of order $e^{\Delta(y)} \sim 10^{-3}$.  Any process that requires the strings to tunnel out of the inflationary well thus involves a ratio of scales that is this large, and pays a penalty of at least $e^{-10^{6}}$ in the Schwinger calculation.  Thus the annihilation is completely suppressed if the image is not in the inflationary throat~\cite{Copeland:2003bj}.  Equivalently, it is suppressed if there is no orientifold fixed point in the throat.  There is no particular reason for the throat to be coincident with a fixed point; their relative positions are fixed by the complex structure moduli, which depend on flux integers, 
and these are expected to take rather generic values~\cite{Bousso:2000xa,Douglas:2003um,Ashok:2003gk}.

The strings can also break on a brane.  The model must include branes on which the Standard Model (SM) fields live.  If these are D3-branes in the inflationary throat the strings will break; if they are D7-branes that pass through the inflationary throat then all but the D1-string will break~\cite{Copeland:2003bj}.  If they are outside the throat then the strings are stable for the same reason as above.  In the simplest implementation of the RS idea the SM branes must be in a different throat: the depth of the inflationary throat is something of order the GUT scale, while the depth of the SM throat should be of order the weak scale.  One must ask whether other branes might still be in the inflationary throat, but this is not possible: these would have low energy degrees of freedom which would receive most of the energy during reheating, rather than the SM fields.  So this gives a simple scenario in which the strings are stable, but much more work is needed to see whether it is viable, and whether it is generic.

The stabilization of moduli has not been as fully developed in models based on large compact dimensions, but many of the same considerations apply.  The warp factor will depend at least weakly on the compact dimensions, because of the symmetry breakings required in realistic models, and this will localize the various branes and strings.  The annihilation of objects that are physically separated in the compact directions is then suppressed due to their separation.

In summary, it is encouraging to see that strings can be stabilized as a side effect of certain generic properties such as warping and/or large dimensions, which are needed to lower the inflationary scale below the Planck scale in these models.  Incidentally, there are examples of exactly stable discrete strings, for example by wrapping higher-dimensional branes on torsion cycles, but it is not clear how a network of these strings would be produced.  It is not sufficient to have one kind of string that is produced, and a different kind that is stable!\footnote{In this connection, the possible stable strings discussed in ref.~\cite{Leblond:2004uc} involve additional degrees of freedom beyond those involved in brane annihilation, and so would not be produced in brane inflation.}

\section{Seeing Cosmic Strings}

We will focus on strings that have only gravitational interactions.  Light matter fields live on branes; in most cases stability requires that the branes and strings be physically separated so the light fields will interact with the strings only through bulk interactions.  In particular there will then be no charged zero modes, as there are in superconducting strings~\cite{Witten:1984eb}.  We have noted at least one way that stable strings can coincide with a brane, namely D1-branes with a D7-brane, but the zero modes will become massive due to symmetry breaking effects unless protected by a chiral symmetry.  It seems difficult to obtain such a symmetry in the models that we are considering; it is an interesting question whether there are models in which stable superconducting strings are produced.  In the absence of chiral symmetries (noting that supersymmetry must ultimately be broken), the {\it only} light or zero modes on cosmic superstrings will be their bosonic coordinates in the noncompact directions.

Thus we must ask what is the likely range for $G\mu$.  In models with large compact dimensions~\cite{Arkani-Hamed:1998rs,Antoniadis:1998ig} this is suppressed by some power of $R/L_{\rm P}$.  In models with large warping~\cite{Randall:1999ee} it is suppressed by $e^{2\Delta}$ as in eq.~(\ref{eq:tenred}).  Thus the tension is essentially a free parameter; for example in the warped models it is the exponential of a ratio of flux quanta~\cite{Giddings:2001yu}.  In models of brane inflation, the value of $G\mu$ can be deduced from the observed value of the CMB fluctuations $\delta T/T$.  That is, one assumes that $\delta T/T$ arises from the quantum fluctuations of the inflaton; this is natural given the flat form of the inflaton potential.  For any given brane geometry the inflaton potential has a definite functional form.  For example, in the D3/$\OL{\rm D3}$ system it is $V \sim V_0 - O(\phi^{-4})$.  Fitting the observed $\delta T/T$ then determines the normalization of $V_0$, and this in turn determines the string tension.  For example, in the \klmt model $V_0$ is the Planck scale times $e^{4\Delta}$ and the string tension is just the square root of this (times $g_{\rm s}^{ 1/2}$ for F-strings and $g_{\rm s}^{ -1/2}$ for D-strings).  This puts the tensions in the likely range $10^{-10} \lsim G\mu \lsim 10^{-9}$~\cite{Kachru:2003sx}.  For models based on large dimensions, for various geometries, refs.~\cite{Sarangi:2002yt,Jones:2003da} find values in the range $10^{-12} \lsim G\mu \lsim 10^{-6}$.

The CMB and pulsar bounds on $G\mu$ quoted in the review are at the upper end of the brane inflation range, ruling out the highest-tension models.  Both bounds will improve in the coming decade, at least by one or two orders of magnitude, due to improved data.  However, the more exciting prospect comes from LIGO~\cite{Damour:2000wa,Damour:2001bk}.  Under most circumstances LIGO is at a disadvantage looking for cosmological backgrounds because these fall with increasing frequency: LIGO is looking at frequencies that are $10^{10}$ times those of the pulsar measurements (100 Hz versus years$^{-1}$).  However, something unexpectedly nice happens.  When a loop of string in three space dimensions oscillates, typically it forms a cusp several times per oscillation~\cite{Turok:1984cn}.  The instantaneous shape is $y = |x|^{2/3}$, with the tip moving at the speed of light in the $x$-direction.  Like the crack of a whip, a great deal of energy is concentrated in the tip, but this whip is perhaps hundreds of light-years long, with tension not so far below the Planck scale, and so it emits an intense beam of gravitational waves in the direction of its motion~\cite{Damour:2000wa}.
The Fourier transform of such a singularity is much larger at high frequency than for a smooth function, large enough that it is within reach of LIGO.  

This is shown in figure~3, reproduced from~\cite{Damour:2001bk}. 
\begin{figure}[t]
\begin{center}
\includegraphics[height=.3\textheight]{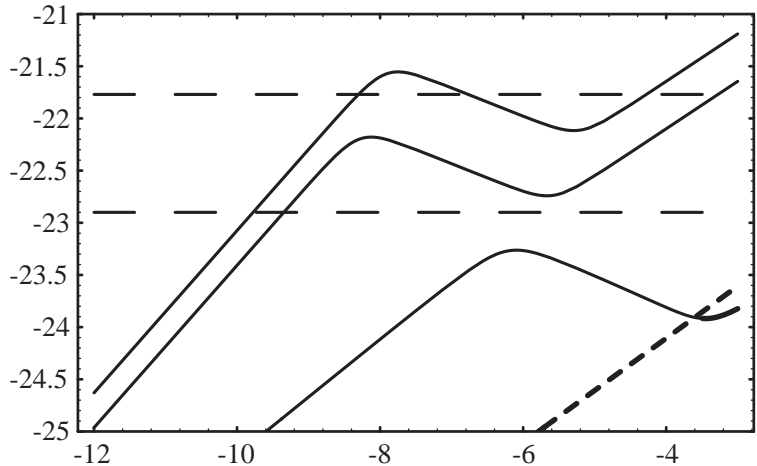}
\end{center}
\caption{Gravitational wave cusp signals, taken from Damour and Vilenkin~\cite{Damour:2001bk}.
The horizontal axis is $\log_{10} \alpha$ where $\alpha = 50G\mu$.  Thus the brane inflation range $10^{-12} \lsim G\mu \lsim 10^{-6}$ becomes $-10.3 < \log_{10} \alpha < -4.3$. The vertical axis is $\log_{10} h$ where $h$ is the gravitational strain in the LIGO frequency band.  The upper and lower dashed horizontals are the sensitivities of LIGO I and Advanced LIGO at one event per year.  The upper two curves are the cusp signal under optimistic and pessimistic network assumptions.  The lowest solid curve is the signal from kinks, which form whenever strings reconnect.  The dashed curve is the stochastic signal.}
\end{figure}
 Under optimistic assumptions (but not, I think, too optimistic), even LIGO I is close to discovery sensitivity of one event per year over much of the range of interesting tensions, including the narrow range of the \klmt model.  This is remarkable: cosmic superstrings might be the brightest objects in gravitational wave astronomy, and the first discovered!  LIGO I to date has around $0.1$ design-year of data, but it is supposed to begin a new science run in January 2005 at close to design sensitivity and with a good duty cycle.  Advanced LIGO is sensitive over almost the whole range, and with a higher event rate.  It has not yet been funded, but it is likely to be in operation around five years from now.  LISA, which may follow a few years after that, is even more sensitive.  In magnitude of $h$ it is comparable to LIGO I, but it is looking at a frequency 10,000 times lower and so the typical strains are 1000 times greater~\cite{Damour:2001bk}.  The cusp events might be seen in a search for unmodeled bursts.  The shape is not as complex as for stellar and black hole inspirals, but modeling the specific frequency dependence will increase the signal-to-noise ratio.  The power-law frequency dependence of the cusp is distinctive.  See also ref.~\cite{Siemens:2003ra} for a discussion of the form of the cusp signal.
 
The dependence of the sensitivity on string tension in figure~3 is interesting because it is not monotonic.  This comes about as follows~\cite{Damour:2001bk}.  As the string tension decreases, the coupling to gravity becomes weaker and so does the gravitational wave burst from a given cusp.  However, since gravitational radiation is the only decay channel for string loops they will live longer.\footnote{There is another effect as well, the dependence of the short-distance network structure on the gravitational radiation, but we can overlook this for simplicity.}  Thus as we decrease $G\mu$ there are more but smaller cusps.  Seeing smaller intrinsic events requires that the events be closer to us.  Thus, the three regimes that are evident in figure~3 correspond at the smallest tensions to cusps that took place at redshifts less than one, at the intermediate tensions to cusps that are at redshifts greater than one but in the matter-dominated era, and at the largest tensions to cusps that occurred during the radiation-dominated era.  The rise in the event rate with decreasing tension in the middle range comes about because the signals from smaller, later, cusps suffer less from redshifting during the relatively rapid matter-dominated expansion.
The pulsar bounds are also strengthed by taking into account the cusps.  This requires careful treatment of statistics, but these might also reach most of the interesting range of tensions~\cite{Damour:2001bk, Damour:2004kw}.

\section{Distinguishing Superstrings}

Let us now imagine the best case, that LIGO has observed some cusp events.  Can we hope to distinguish a network of F- and/or D-strings from a network of gauge theory soliton strings?

The microscopic structure of the string core does not affect the evolution of strings that are light-years in length, except when two strings cross and their cores interact.  We have noted that gauge theory solitons will always reconnect.  For F-strings, reconnection is a quantum process, and takes place with a probability $P$ of order $g_{\rm s}^2$.  The numerical factors are worked out in ref.~\cite{Jackson:2004zg}.  To be precise, $P$ is a function of the relative angle and velocity in the collision, but it is simplest to the value averaged over collision parameters.  

An important issue is the motion of the string in the compact dimensions.   For many supersymmetric compactifications, strings can wander over the whole compact space.  Thus they can miss each other, leading to a substantial suppression of $P$~\cite{Jones:2003da,Dvali:2003zj}.  However, we have noted that in realistic compactifications strings will always be confined by a potential in the compact dimensions.  Even if the scale of the potential is low, the fluctuations of the strings are only logarithmic in the ratio of scales (this is characteristic of one-dimensional objects)~\cite{Copeland:2003bj}.  Thus there is no suppression by powers of the size of the compact dimensions, but the logarithm can be numerically important --- it tends to offset powers of $\pi$ that appear in the numerator.  The value of $g_{\rm s}$, and the scale of the confining potential, are not known, but in a variety of models ref.~\cite{Jackson:2004zg} finds $10^{-3} \lsim P \lsim 1$.  For D-D collisions the situation is more complicated, and in the same models one finds $10^{-1} \lsim P \lsim 1$.  For F-D collisions, $P$ can vary from 0 to 1.

Given the value of $P$, to determine the observational effect one must feed this into the network simulations.  A simple argument suggests that the signatures scale as $1/P$: the amount of string in the network must be increased by this factor in order for an increased number of collisions (per unit length of string) to offset the reduced $P$ in each collision~\cite{Damour:2004kw}.  This is a bit oversimplified, because there are issues connected with the sub-horizon scale structure in the string network~\cite{Bennett:1990uz, Austin:1993rg} that can work in either direction.  

 If we take $1/P$ as a model, we see that for the smaller values of $P$ discussed above there can be a substantial increase in the signal even above the encouraging values found in the last section, so that LIGO might soon begin to see {\it many} cusps.  Of course, the existing bounds become stronger, e.g. $G\mu < 10^{-7}/P$ from pulsars.  If $P$ is only slightly less than one, say $0.5$, then it will require precision simulation of the networks and good statistics on the signatures to distinguish this from 1.0.  It should be noted that even with given values of $\mu$ and $P$ there are still substantial uncertainties in the understanding of the behavior of string networks.  This has recently been discussed in ref.~\cite{Damour:2004kw}, which concludes that the sensitivities given in figure 3 are only weakly dependent on the unknowns.
 
 To first approximation there are two relevant parameters, $\mu$ and $P$.  Each individual cusp event has only a single parameter to measure, its overall strength $h$: because it is a power law  there is no characteristic frequency scale.  (There is a high frequency cutoff, determined by the alignment of the cusp with the detector~\cite{Damour:2001bk}, but this gives no information about the cusp itself.)  After $O(10)$ cusps are seen one can begin to plot a spectrum, $dN \sim A h^{-B} dh$, and from the two parameters $A$ and $B$ fix $\mu$ and $P$.
There are degeneracies --- $B$ depends primarily on the epoch in which the cusp took place --- but with a more detailed spectrum, and ultimately with data from kink events and pulsars, this degeneracy will be resolved.  Thus $\mu$ and $P$ will be overdetermined, and nonstandard network behavior (such as we are about to discuss) will be detectable.

The second potentially distinguishing feature of the superstring networks is the existence of both F- and D-strings~\cite{Copeland:2003bj, Dvali:2003zj}, and moreover bound states of $p$ F-strings and $q$ D-strings with a distinctive tension formula
\begin{equation}
\mu = \mu_0 \sqrt{p^2 + q^2/g_{\rm s}^2}\ . \label{eq:tensions}
\end{equation}
In this case, when strings of different types collide, rather than reconnecting they form more complicated networks with trilinear vertices.  It is then possible that the network does not scale, but gets into a frozen phase where it just stretches with the expansion of the universe~\cite{Kibble:1976sj,Vilenkin:1984rt}.  If so, its density would come to dominate at the tensions that we are considering.  The F-D networks have not yet been simulated, but simulations of comparable networks suggest that they scale, possibly with an enhanced density of strings~\cite{Vachaspati:1986cc,Spergel:1996ai,McGraw:1997nx}.
From the discussion above, it follows that one will not directly read off the spectrum~(\ref{eq:tensions}) from the observations, but there should eventually be enough information to distinguish F-D string networks from other types.

Networks with multiple types of string can also arise in field theories, though I do not know any {classical} field theory that gives the particular spectrum~(\ref{eq:tensions}).  However, because of duality there will be gauge theory strings that are very hard to differentiate from the F- and D-strings that we are discussing.  In particular, in the \klmt model the strings exist in a Klebanov-Strassler~\cite{Klebanov:2000hb} throat which has a dual description as a cascading gauge theory.\footnote{For discussions of relations between field theory strings and F- and D-strings see~\cite{Becker:1995sp,Edelstein:1995ba,Dvali:2003zh,Halyo:2003uu,Binetruy:2004hh,Gubser:2004qj,Achucarro:2004ry,Gubser:2004tf,Lawrence:2004sm}; see~\cite{Rocher:2004uv,Rocher:2004my}.}
  Thus the $(p,q)$ spectrum, and well as the property $P \ll 1$, do arise in this theory.  The point is that the dual description of the F-strings is as electric flux tubes, which are quantum mechanical objects rather than classical solitons, whereas almost all work on field theory cosmic strings has been in the context of perturbative unification.  The possibility that electric flux tubes could be cosmic strings was discussed in~\cite{Witten:1985fp}.  
  
Indeed, the existence of dualities between string theories and field theories raises the issue, what really is string theory?  This is beyond our current scope, but I note that in the present case there is a quantitative question.  The \klmt model has a paremeter $gM$; when this is large the string description is the valid one, and when it is small the gauge description is the valid one.  To fantasize about the maximum possible information that might be extracted from the string network, let us suppose that we could map it out in detail by lensing.  In this case we could see the spectrum~(\ref{eq:tensions}) in the deficit angles, and by measuring various correlations in the network we might deduce the actual function~$P(v,\theta)$.  This is obtained as a piece of the Virasoro-Shapiro amplitude~\cite{Polchinski:1988cn, Jackson:2004zg}, and so in this best case we might truly see string theory written in the sky.

\section{Conclusions}

As we have seen, each of the four conditions that we discussed at the beginning is independently model-dependent.\footnote{I think that I am presenting this in an overly pessimistic way.  When I hear talks on new physics (especially in the context of cosmology!) there is much less of a sense of an a priori measure of what is likely.  I am quite sure that many of the ideas that are prominently discussed have an a priori probability very much less than cosmic superstrings. (I had started to include a list, but decided that it would be distracting and inflammatory.  It was rather long.)}
However, quite a number of things have worked out surprisingly well: the production of strings in brane inflation, the possible stabilization of the strings as a side effect of other properties of the models (in particular, of the stabilization of the vacuum itself), the possibility to see strings over many interesting orders of magnitude of tension, and the existence of properties that distinguish different kinds of string so that after the strings are discovered we can do a lot of science with them.  In any case, searching for cosmic strings is a tiny marginal cost on top of experiments that will already be done, and it is great that string theorists will have a stake in these experiments over the coming decade or more.

\section{Acknowledgments}

I would like to thank E. Copeland, M. Jackson, N. Jones, and R. Myers for collaborations, and  N. Arkani-Hamed, L. Bildsten, G. Dvali, A. Filippenko, T. Kibble, A. Lo, A. Lommen, 
J. Preskill, H. Tye, T. Vachaspati, and A. Vilenkin for discussions and communications.  This work was supported by National Science Foundation
grants PHY99-07949 and PHY00-98395.

\end{document}